\documentclass[sigconf]{acmart}

\AtBeginDocument{%
  }
\copyrightyear{2025}
\acmYear{2025}
\setcopyright{acmlicensed}\acmConference[UIST '25]{The 38th Annual ACM Symposium on User Interface Software and Technology}{September 28-October 1, 2025}{Busan, Republic of Korea}
\acmBooktitle{The 38th Annual ACM Symposium on User Interface Software and Technology (UIST '25), September 28-October 1, 2025, Busan, Republic of Korea}
\acmDOI{10.1145/3746059.3747613}
\acmISBN{979-8-4007-2037-6/2025/09}

\newcommand{\note}{Rhythm Note}
\newcommand{\group}{Rhythm Group}
\newcommand{\view}{Rhythm Waterfall}
\newcommand{\system}{\textit{RhythmTA} }

\usepackage{xcolor}
\usepackage{soul}    % 提供高亮命令
\colorlet{myhlcolor}{yellow!30}  % 自定义高亮颜色（可选）
\sethlcolor{myhlcolor} 

\usepackage{enumitem}
\usepackage[ruled,vlined]{algorithm2e}
\usepackage{amsmath, bm}
\definecolor{revision_color}{RGB}{0,0,0}
\definecolor{mygreen}{RGB}{0,176,80}

\usepackage{hyperref}

\begin{document}

%%
%% The "title" command has an optional parameter,
%% allowing the author to define a "short title" to be used in page headers.
\title{RhythmTA: A Visual-Aided Interactive System for ESL Rhythm Training via Dubbing Practice}

%%
%% The "author" command and its associated commands are used to define
%% the authors and their affiliations.
%% Of note is the shared affiliation of the first two authors, and the
%% "authornote" and "authornotemark" commands
%% used to denote shared contribution to the research.
\author{Chang Chen}
\email{cchenda@connect.ust.hk}
\affiliation{%
  \institution{Hong Kong University of Science and Technology}
  \city{Hong Kong SAR}
  \country{China}
}

\author{Sicheng Song}
\email{csescsong@ust.hk}
\affiliation{%
  \institution{Hong Kong University of Science and Technology}
  \city{Hong Kong SAR}
  \country{China}
}

\author{Shuchang Xu}
\email{sxuby@connect.ust.hk}
\affiliation{%
  \institution{Hong Kong University of Science and Technology}
  \city{Hong Kong SAR}
  \country{China}
}

\author{Zhicheng Li}
\email{zlijw@connect.ust.hk}
\affiliation{%
  \institution{Hong Kong University of Science and Technology}
  \city{Hong Kong SAR}
  \country{China}
}

\author{Huamin Qu}
\email{huamin@cse.ust.hk}
\affiliation{%
  \institution{Hong Kong University of Science and Technology}
  \city{Hong Kong SAR}
  \country{China}
}

\author{Yanna Lin}
\authornote{The corresponding author.}
\email{ylindg@connect.ust.hk}
\affiliation{%
  \institution{Hong Kong University of Science and Technology}
  \city{Hong Kong SAR}
  \country{China}
}

\begin{abstract}
English speech rhythm, the temporal patterns of stressed syllables, is essential for English as a second language (ESL) learners to produce natural-sounding and comprehensible speech. Rhythm training is generally based on imitation of native speech. However, it relies heavily on external instructor feedback, preventing ESL learners from independent practice. To address this gap, we present \textit{RhythmTA}, an interactive system for ESL learners to practice speech rhythm independently via dubbing, an imitation-based approach. The system automatically extracts rhythm from any English speech and introduces novel visual designs to support three stages of dubbing practice: (1) \textit{Synchronized \textbf{listening} with visual aids to enhance perception}, (2) \textit{Guided \textbf{repeating} by visual cues for self-adjustment}, and (3) \textit{Comparative \textbf{reflection} from a parallel view for self-monitoring}. Our design is informed by a formative study with nine spoken English instructors, which identified current practices and challenges. A user study with twelve ESL learners demonstrates that \textit{RhythmTA} effectively enhances learners’ rhythm perception and shows significant potential for improving rhythm production.
\end{abstract}

%%
%% By default, the full list of authors will be used in the page
%% headers. Often, this list is too long, and will overlap
%% other information printed in the page headers. This command allows
%% the author to define a more concise list
%% of authors' names for this purpose.
\renewcommand{\shortauthors}{C. Chen, S. Song, S. Xu, Z. Li, H. Qu, and Y. Lin}
%%
%% Article type: Research, Review, Discussion, Invited or position
\acmArticleType{Review}
%%
%% Links to code and data
% \acmCodeLink{}
% \acmDataLink{}
%%
%% Authors' contribution
% \acmContributions{BT and GKMT designed the study; LT, VB, and AP conducted the experiments, BR, HC, CP and JS analyzed the results, JPK developed analytical predictions, all authors participated in writing the manuscript.}
%%
%% Sometimes the addresses are too long to fit on the page.  In this
%% case uncomment the lines below and fill them accodingly.
%%
%% \authorsaddresses{Corresponding author: Ben Trovato,
%% \href{mailto:trovato@corporation.com}{trovato@corporation.com};
%% Institute for Clarity in Documentation, P.O. Box 1212, Dublin,
%% Ohio, USA, 43017-6221}
%%
%%
%% Keywords. The author(s) should pick words that accurately describe
%% the work being presented. Separate the keywords with commas.

\begin{teaserfigure}
  \includegraphics[width=\textwidth]{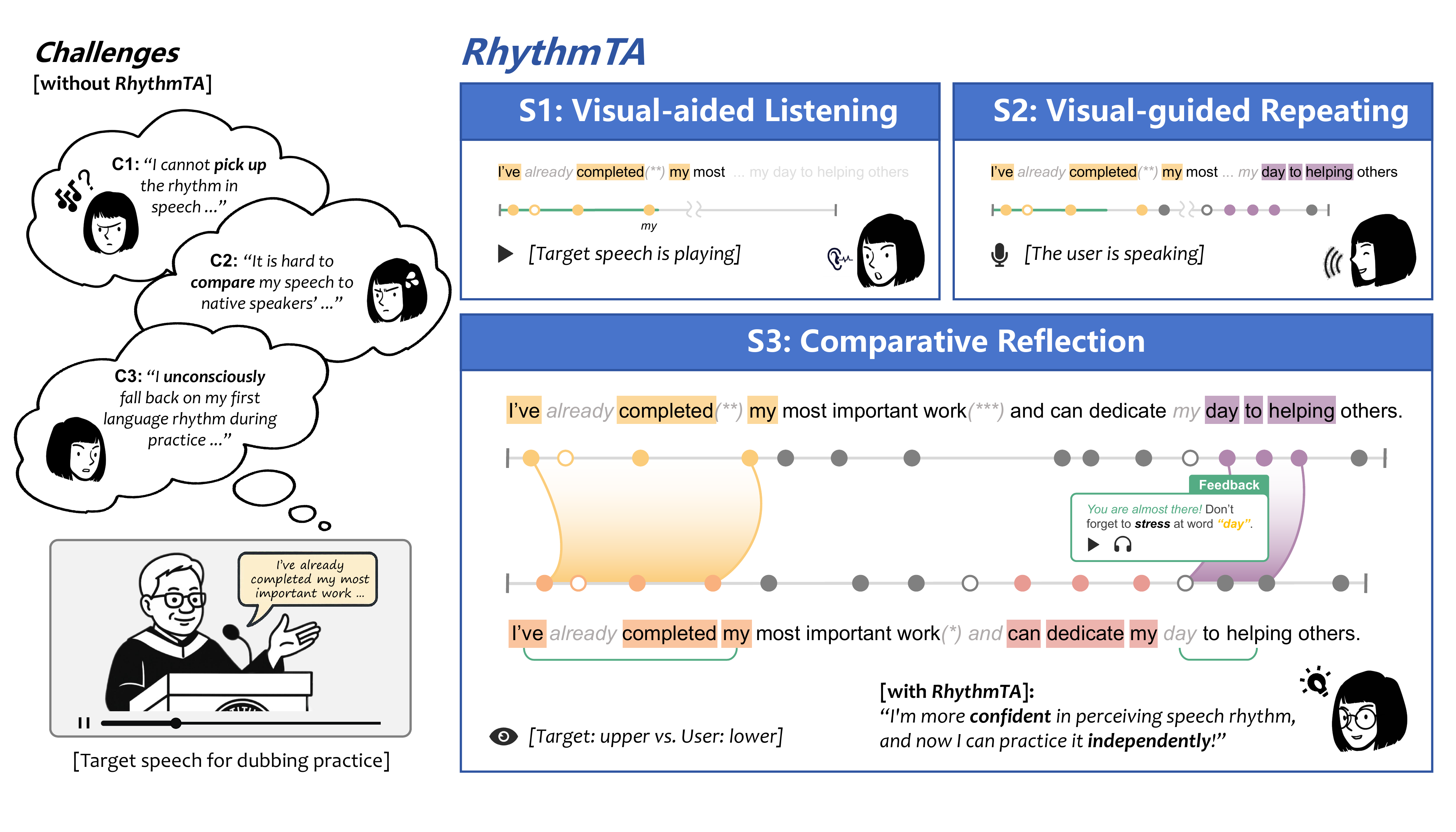}
  \caption{\system helps ESL learners practice English speech rhythm by providing visual aids across three stages—visual-aided listening, visual-guided repeating, and comparative reflection. These designs address key challenges of audio-only rhythm learning, including difficulty in perceiving rhythm (C1), comparing one’s own speech to native speech (C2), and unconsciously reverting to first language (L1)  rhythmic patterns (C3).}
  \label{fig:teaser}
\end{teaserfigure}

\maketitle

\section{Introduction}
Speech rhythm serves to signal syntactic boundaries, convey pragmatic intents~\cite{degano2024speech, roncaglia2013speech}, and enhance communication between individuals. 
English speech rhythm is characterized by alternating stressed and unstressed syllables~\cite{nolan2014speech, CouperKuhlen1986AnIT}, while other spoken languages also exhibit distinct rhythmic patterns~\cite{roach1982distinction, dauer1983stress}.
These rhythmic differences often pose challenges for English as a second language (ESL) learners, affecting their ability to produce natural-sounding speech and potentially hindering communicative comprehension~\cite{peelle2012neural, fujii2014role}. 
% enhancing communication between individuals (Hawkins, 2014; Kotz et al., 2018). 
For ESL learners, acquiring appropriate English rhythmic patterns improves both the naturalness and comprehensibility of their spoken English~\cite{nishihara2014rhythm}.

% However, learning to master English rhythm remains challenging for ESL learners.
Despite extensive studies on approaches to English learning in areas such as pronunciation and voice modulation skills (e.g., pitch, volume, and speed), specific process and challenges of rhythm training remain largely underexplored. 
Traditional spoken English training often follows an imitation-based approach, such as shadowing and dubbing, where learners are exposed to native speech through extensive listening and repeated imitation.
One common pain point is that learners require guidance and feedback from expert instructors, since they often struggle to identify how their speech deviates from the target speech.
% One common pain point is that, while abundant authentic English corpus avaliable online for imitation, learners often struggle to identify how their speech deviates from the target speech.
% Learners still require guidance and feedback from expert instructors, which are not always available and accessible at scale.
To reduce reliance on instructors and facilitate independent learning, several studies~\cite{zhang2020withyou, reza2021designing, wang2020voicecoach} and commercial applications, such as Lingodub~\cite{Lingodub} and Liulishuo~\cite{Liulishuo}, provide feedback on pronunciation, fluency, or voice modulation skills.
However, these systems fail to adapt to rhythm training directly since they neglect the critical components of English rhythm, such as stress and its timing.
This gap restricts ESL learners' ability to independently practice with awareness of rhythmic deviations, impeding their progress in mastering speech rhythm.
% Despite the extensive studies in HCI and media have investigated approaches to English learning, they focus on the aspects like pronunciation, voice modulation skills (such as pitch, volume, and speed), and xxxx.
% The unique features of rhythm (i.e., the combination of stress plancement and timing) arise specific challenges in practicing rhythm.

% Imitation-based pedagogy (e.g., dubbing and shadowing) has proven effective for improving speech rhythm~\cite{}. Many studies support the use of shadowing~\cite{zhang2020withyou, reza2021designing}, and commercial applications for dubbing practice are also available~\cite{Lingodub, Liulishuo, Mofunshow}. 
% However, these tools typically provide feedback only on pronunciation or fluency, leaving ESL learners without automated guidance on speech rhythm.
% This gap restricts their ability to independently practice with awareness of rhythmic deviations, impeding their progress in mastering speech rhythm.

To address this gap, we conducted a formative study with three English education experts and six English speaking tutors to understand the current practices and specific challenges ESL learners encounter in practicing speech rhythm.
The study revealed three key challenges: First, many ESL learners, especially those less sensitive to auditory rhythm discrimination, \textbf{struggle to perceive rhythmic patterns in speech} during listening.
% , hindering imitation.
Second, ESL learners often \textbf{find it difficult to compare their speech to target models} and identify rhythmic flaws.
It prevents them from monitoring their rhythm performance and reflecting for correction.
Third, as native rhythmic patterns are usually resistant to modification, ESL learners often \textbf{unconsciously follow the first language (L1) rhythmic patterns} instead of making intentional adjustment during practice. 
Based on these findings, we summarized six design requirements for a facilitation system.

Informed by the formative study, 
we present \textit{RhythmTA}, a visual-aided interactive system designed for ESL learners’ rhythm training. The system incorporates video dubbing practice, aiming to minimize the discrepancy between a target speech and the user’s utterance. Specifically, \system features an automated rhythm extraction pipeline that detects and extracts rhythm in the form of stress timing from any given English speech. 
Leveraging these rhythmic attributes, \system further introduces a set of intuitive visual designs to support the three stages (as shown in \autoref{fig:teaser}) of dubbing practice:
\textbf{(1) Visual-aided listening}: 
During the listening stage, ESL learners perceive rhythm through visual aids that display the current stress timing in sync with the target speech’s audio playback.
\textbf{(2) Visual-guided repeating}: 
In the repeating stage, ESL learners practice dubbing with the guidance of static visual cues that depict the complete stress timing of the target speech.
\textbf{(3) Comparative reflection}: 
During the reflection stage, ESL learners use a parallel view to compare their rhythm with the target speech. Additionally, in-situ corrective feedback and local replay controls help users identify mistakes.

A user study with twelve participants reported an effective rhythm improvement and a positive learning experience, with an updated understanding of English speech rhythm when practicing dubbing with \textit{RhythmTA}.

% no facilitation application have been found for rhythm training specific. ESL learners cannot access to automatic feedback, which may [阻碍] them from practicing speech rhythm independently.

To sum up, our contributions are three-fold:
% \begin{compactitem}
\begin{itemize}[noitemsep,topsep=0pt,parsep=0pt,partopsep=0pt]
    \item We reveal three challenges faced by ESL learners in improving speech rhythm from a formative study (N=9) with both English education experts and English speaking tutors.
    \item We introduce \textit{RhythmTA}, an interactive system featuring automated rhythm extraction and visual aids to support ESL learners through listening, repeating, and reflection stages during dubbing practice. 
    % \textsc{red{}An auditory-to-visual pipeline is embedded to extract the rhythm from any given English speech.}
    \item We conduct an evaluation study (N=12) to assess system usability, \textcolor{revision_color}{learning facilitation}, \textcolor{revision_color}{learning improvement}, and learning experience of \textit{RhythmTA}.
% \end{compactitem}
\end{itemize}

\section{Related Work}
\subsection{Detecting Stress for Rhythmic Patterns}
Stress and rhythm determine a natural-sounding pronunciation of segments in English~\cite{sole1991stress}.
Previously, linguists ~\cite{pike1945intonation, abercrombie2019elements} believed that English is a stress-timed language with relatively equal durations between successive stressed syllables, thus establishing a hypothesis called isochrony. Although empirical evidence for strict isochrony has not been found, and later experts held an opposite opinion ~\cite{dauer1983stress}, there is still consensus that the rhythmic beats were closely associated with the stressed syllables in English ~\cite{allen1972location}. 

Native English speakers tend to stress via longer durations, higher pitches, or volumes ~\cite{fokes1984patterns} to make it prominent in utterances. There are lexical stress, located at a syllable within a word, and sentence stress, which is the prominence of the word in a sentence. Both can be detected from speech given the corresponding datasets. Prior work on lexical stress detection~\cite{xia2019attention, ruan2019end, mallela2023comparison, mallela2024comparative} utilized deep learning approaches to classify the syllable segments in a word as primary stress, secondary stress, and non-stress. To capture the speech rhythm, however, a sentence stress detection model is required instead. 

In prior works, Lee et al. ~\cite{lee2017automatic} achieved sentence stress detection based on handcrafted features. Lin et al. ~\cite{lin2021improving} utilized a bi-directional LSTM and relied on phoneme-level segmentation to predict sentence stress. As wav2vec 2.0~\cite{baevski2020wav2vec} has been proposed, it works as a powerful speech representation model in recent years with proven sensitivity to stress~\cite{bentum2024processing}. In this work, we adopt wav2vec 2.0 to extract acoustic features from word audio segments and build a Conformer-based~\cite{gulati2020conformer} sentence stress detection model. The training dataset comes from Aix-MARSEC corpus~\cite{auran2004aix}.
While current speech technology is capable of stress detection, neither interactive learning systems nor intuitive visual representations specific to rhythm have been proposed yet. This leaves a gap for ESL learners in understanding this intricate yet significant concept.

\subsection{Enhancing English Speaking via Imitation}
Imitation practice plays an important role in spoken language learning~\cite{jones1997beyond}. ESL learners benefit from imitation practice by exposing in authentic English speech and mimicking native pronunciation, intonation, and rhythm at the same time~\cite{tepperman2010testing}. 

One common practice is shadowing, which requires learners to utter the exact words immediately once they hear them. It allows little time for learners to think but just reproducing the speech relatively subconsciously. The HCI community has made several in-depth research on facilitating ESL learners in shadowing practice, such as WithYou~\cite{zhang2020withyou}, and Designing CAST~\cite{reza2021designing}.
While this approach \textcolor{revision_color}{contributes to} both listening and speaking improvement~\cite{hamada2012effective}, it \textcolor{revision_color}{imposes} cognitive load due to its narrow window between listening and repeating.

Dubbing is another popular approach for imitation-based speaking practice. It requires ESL learners to revoice video clips from English movies, TV dramas, or  public speeches, mimicking the original speech as closely as possible after listening to it freely. This helps learners reproduce utterances of more authentic prosody and enhanced naturalness~\cite{luo2016naturalness}. There are multiple commercial applications designed for English learners to practice video dubbing, such as 
Lingodub~\cite{Lingodub}, Liulishuo~\cite{Liulishuo}, and Mofunshow~\cite{Mofunshow}. 

Although these apps assess certain useful aspects, such as pronunciation and fluency, none of them can explicitly detect prosodic features nor guide users in improving speech rhythm.
ESL learners who are less sensitive to critical rhythmic features (e.g., stress, pausing, and rate) may struggle to perceive and reproduce the original speech's rhythm. Without external feedback on rhythm, they are less likely to make improvements.

\subsection{Visualizing Prosodic Features in Speech}
Sophisticated visualizations of prosodic features have been continuously proposed. 
Pitch, volume, pausing, and stress are the most basic elements to visualize~\cite{rosenberger1999prosodic, patel2011readn, yuan2019speechlens, wang2020voicecoach, chaudhary2021verbose}.
They intuitively demonstrate the auditory information of speech, facilitating the goal of language learning or public speaking training.

The visualizations of prosodic features have been achieved by manipulating the font ~\cite{rosenberger1999prosodic, patel2011readn, chaudhary2021verbose}, making markings in addition to the text~\cite{wang2020voicecoach, rubin2015capture}, augmenting directly in the background of the script~\cite{patel2011readn, oktem2017prosograph, yuan2019speechlens}, or using threaded wave-forms directly~\cite{yoon2014richreview, venkataramani2017autodub}.
Most of these works reserved neat typesetting of the script text, while ReadN’Karaoke~\cite{patel2011readn} selected to rearrange the words horizontally according to their uttered timestamps. Both designs can clearly associate a variable of any acoustic features mentioned above with a local visual attribute (e.g., height, color, or font weight); however, speech rhythm is different. As it represents regularity in the timing of consecutive speech units, the visualization of rhythm should consider a contextual range of data, extracting the pattern within instead of displaying raw variables. 

Musical notation provides a classic way to visualize rhythm using divisible time units~\cite{liu2012mathematical}, such as whole notes, half notes, quarter notes, and finer subdivisions of notes and rests. However, speech instead does not exhibit such regular units all the time, failing to fit into a music sheet elegantly. Approaches to visualizing temporal patterns based on rather random signals have also been researched. 
Begole et al. ~\cite{begole2003rhythm} modeled the rhythm of staff availability from their online presence data, providing an example of rhythm visualization through the use of the color saturation gradient.

In this work, we start with speech stress detection, compute the intervals between stressed words, and visualize their temporary stability and overall fluctuations via a novel design of rhythm groups. It serves as an intuitive representation of speech rhythm, complementing the nuanced and complex auditory information through the visual channel.

\section{FORMATIVE STUDY}

We conducted a formative study to identify the current practices and challenges in the rhythm training process for ESL learners, and to derive design requirements for facilitating rhythm training.
This study involved three experienced English education experts from the university and six private English speaking tutors, all of whom regularly teach speaking lessons.
% with an active lesson schedule.
These experts are knowledgeable about natural English speech rhythm and have extensive experience instructing ESL learners in English speaking.
Rather than focusing on learners directly, we chose to interview instructors to gain a broader perspective grounded in long-term and first-hand teaching experience.
This approach allowed us to uncover recurring learner difficulties that may not be easily articulated by individual students.

\subsection{Study Setup}

\subsubsection*{\textbf{Participants}} 
We recruited three English education experts (E1-3; two females, one male; aged  from 40 to 58) through advertisements distributed via our university email list.
They have an average of $23.7\pm 8.5$ years of teaching experience at the university. 
Two of them have taught courses on English speaking fluency, and one has taught academic English speaking.
Their students are primarily non-native English speakers, including undergraduates and postgraduates.
We also recruited six private English speaking tutors (E4-9; three females, three males; aged from 22 to 70) through online advertisements on social media. 
These tutors work actively with ESL learners, averaging $23.0\pm 8.8$ teaching hours per week.

\subsubsection*{\textbf{Protocol}}
The formative study was conducted through one-on-one online or face-to-face interviews.
We first obtained consent forms from the participants, which authorized us to record audio during the sessions and collect their demographic information and feedback for research purposes.
Each session lasted around one hour, and participants were offered a compensation of US\$14/h.

\subsubsection*{\textbf{Procedure}}
We began each session by collecting participants' demographic information and teaching experience in English speaking.
We then asked them to describe their current practices for training students in English speech rhythm.
Following that, we invited them to discuss the challenges ESL learners face during rhythm training.
We used open-ended questions such as, ``\textit{Did your students encounter any challenges during this training process}'', and ``\textit{What aspects, if any, hindered their rhythm learning}''. 
To better understand these challenges, we followed up by asking for detailed explanations and concrete examples, such as ``\textit{Can you provide specific examples of why and how this aspect hindered students' rhythm training}''.
Finally, we invited participants to propose potential tool features they believed could help ESL learners overcome the identified challenges and support their rhythm learning.

\subsection{Findings}
\label{subsec: challenges}
Although speech rhythm is discussed less frequently than pronunciation and intonation in English speaking courses and it is not necessary for all ESL learners to practice, all participants in the formative study agreed that speech rhythm is crucial for achieving greater naturalness, comprehensibility, and even communicative effectiveness in spoken English. 

Nearly all of them (E1-E7, E9) have intentionally taught ESL learners how to improve speech rhythm. They primarily use an imitation-based approach, typically involving three stages: listening, repeating, and reflection.
In the listening stage, the instructors either deliver a model speech themselves or play English audiovisual clips to learners. 
This provides learners with a concrete example of natural English speech rhythm.
In the repeating stage, learners are encouraged to repeat after the instructors or imitate the model speech. When using movie or TV clips as imitation materials, several instructors (E2, E3, E5, E9) emphasized that learners should strive to mimic not just pronunciation, but also the emotional tone, speech rate, and stress patterns.
In the reflection stage, instructors offer instructions to help learners identify and correct deviations.

While this approach is effective in the classroom and necessary for rhythm improvement, most instructors (E1-E3, E6-E8) reported that the training time was insufficient. E6 and E7 mentioned a common issue among ESL learners: they often forget instructions from previous lessons when attending classes on a weekly basis. \textit{``They come back every week as if starting from scratch!''} (E7). E1 noted: \textit{``Sometimes students just need a coach to help them detect the deviations in their speech, but that’s not available after class.''} Regarding the dilemma between sustained post-class practice and reliance on instructor feedback, instructors identified several challenges that that hinder learners' independent speech rhythm practice.

\begin{figure*}[t]
    \centering
    \includegraphics[width=1\linewidth]{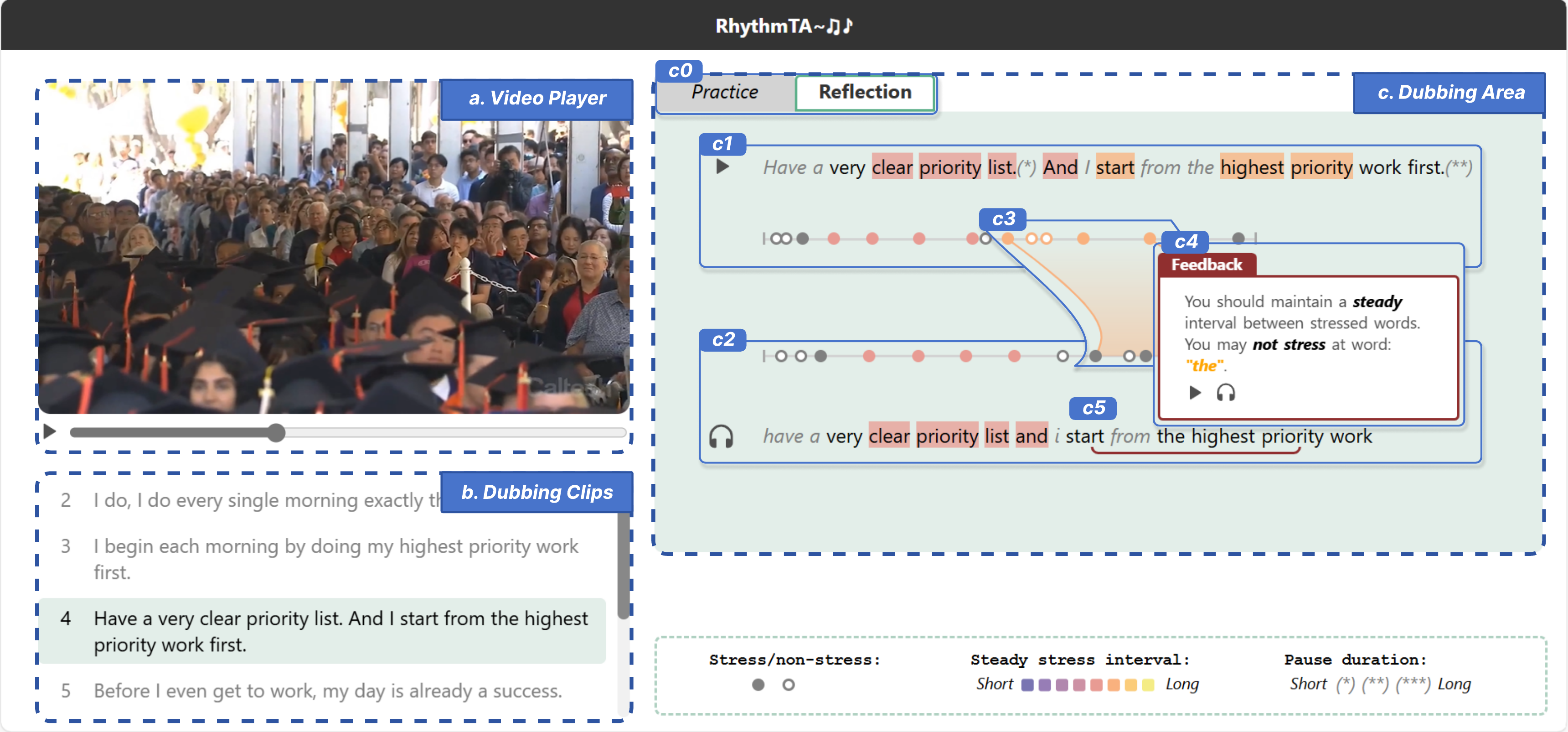}
    \caption{The \system interface includes  three main components: (a) the video player for viewing the original content, (b) the dubbing clips panel displaying text segments for practice, and (c) the dubbing area integrating visual and interactive feedback across three learning stages: listening, repeating, and reflection. Within the dubbing area, c0 suggests the current learning stage, where \textit{``Practice''} includes listening and repeating, while \textit{``Reflection''} indicates the reflection stage. c1 displays the target rhythm notation, visualizing stress and stress timing along a timeline. After completing a dubbing attempt, comparative components c2-c5 become active: c2 shows the rhythm notation of user speech, c3 highlights areas of target rhythm groups for intuitive comparisons between the target and user speech, c4 provides corrective feedback with contextual audio replay, and c5 visually annotates performance using colored curly braces to indicate rhythm accuracy. Together, these components guide users in refining their speech rhythm effectively.}
    \label{fig:interface}
\end{figure*}

\subsubsection*{\textbf{C1: ESL learners struggle to perceive rhythmic patterns when listening to English speech}}
Instructors noted that many ESL learners struggle to detect key rhythmic features from speech, such as stress placement, pausing, and speech rate variations.
E2 shared her observation that students who had learned musical instruments tended to pick up English speech rhythm faster. She acknowledged substantial individual differences in rhythm sensitivity among ESL learners, which appeared to correlate with learning outcomes.
This gap in auditory rhythm discrimination impairs learners' perception of rhythmic patterns and thus their understanding of natural speech rhythm.
Also, it directly hinders learners' ability to evaluate their own speech rhythm and identify rhythmic deviations (E6-E9).

\subsubsection*{\textbf{C2: ESL learners face challenges in comparing their speech rhythm with target speech}}
Instructors mentioned that ESL learners typically assess their own speech by recording and comparing it with a native example.
Even if learners can perceive the rhythmic patterns of each speech\textcolor{revision_color}{,} the linear nature of audio makes this comparison cognitively demanding.
Simultaneous playback causes auditory interference and temporal misalignment, while sequential playback relies heavily on short-term memory.
As a result, learners often miss subtle rhythmic differences and become frustrated when repeatedly switching between the two recordings.  
Instructors pointed out that these limitations prevent learners from effectively identifying where and how their speech deviates from the target rhythm, which further prevents them from making targeted improvements in subsequent exercises.

\subsubsection*{\textbf{C3: ESL learners unconsciously default to the rhythm patterns of their first language (L1)}}
Instructors observed that many ESL learners carry over rhythmic habits from their native languages when speaking English.  
For instance, learners with a Chinese language background often produce syllables with relatively equal duration and intensity~\cite{chen1996new}, in contrast to the alternating stress patterns of English.  
E6 highlighted the most significant rhythmic problems among her students: \textit{``They don’t realize they should pause briefly at the end of a sentence, as English has a very different rhythm compared to their native language. This issue can make their communication lack emphasis or a sense of conclusion.''}
Even when corrective feedback is provided by tutors, ESL learners often struggle to adjust themselves intentionally throughout the whole speech.  
Instructors pointed out that without external support to signal appropriate rhythmic performance in time, such practice drills may have limited effectiveness.

\subsection{Design Requirements}
\label{subsec: design requirement}
Building on the current practices and challenges revealed in our formative study, we derive six design requirements for an interactive rhythm training system for ESL learners to use independently.

\subsubsection*{\textbf{DR1: Follow the imitation-based learning method}}
As instructors noted in the formative study, rhythm training for ESL learners typically uses an imitation-based approach.  
Therefore, the system should follow this method, guiding learners through listening to authentic English speech, repeating the speech, and reflecting on their own performance to facilitate continuous improvement.

\subsubsection*{\textbf{DR2: Enable easy perception of rhythmic pattern in speech}}
The system should provide aids for ESL learners in perceiving rhythmic patterns in any speech (\textbf{C1}).  
With the system, learners with varying phonetic sensitivity should be able to identify critical rhythmic features, such as stress, pausing, and speech rate, and track their temporal dynamics.

\subsubsection*{\textbf{DR3: Provide real-time guidance during dubbing practice}} 
The system should offer external support to ESL learners during dubbing by signaling appropriate rhythmic patterns in real time (\textbf{C3}).  
Such guidance can help learners intentionally adjust their stress, timing, and pacing to follow the rhythmic patterns of the target speech, rather than unconsciously reverting to the prosodic habits of their native language.

\subsubsection*{\textbf{DR4: Support intuitive comparison between self speech and target speech}}
The system should help ESL learners easily compare the rhythm of their own speech with that of the target speech and identify where and how it deviates from native rhythmic patterns (\textbf{C2}).  
Rather than relying solely on linear audio playback, the system should enable learners to perceive and contrast the rhythm of both speech simultaneously.
This can reduce cognitive load and minimize the time and effort spent on repeated switching and sequential replay.

\subsubsection*{\textbf{DR5: Offer corrective feedback to guide subsequent dubbing}}  
Instructors in our formative study typically provided learners with feedback after each practice to help them improve.  
To support similar progress in independent learning scenarios, the system should offer performance-based feedback and actionable guidance for learners to refine their rhythm in the next round of practice.

\subsubsection*{\textbf{DR6: Reduce learners' anxiety during practice}}
Instructors noted that overly strict expectations, such as requiring perfect rhythmic regularity or exact stress timing, can increase learners’ anxiety and reduce their learning motivation.  
The system should adopt a tolerant and flexible approach to rhythm evaluation to reduce stress during practice.

\section{\textit{RHYTHMTA}}

Elicited by the formative study, we adopt one common and engaging imitation-based practice (\textbf{DR1}), \textit{video dubbing}, for independent rhythm training. We create an interactive practice environment with three learning stages in the \system system. Specifically, it consists of \textit{visual-aided listening} (\textbf{DR2}), \textit{visual-guided repeating} (\textbf{DR3}), and \textit{comparative reflection} (\textbf{DR4}). Corrective feedback based on users' rhythm deviations (\textbf{DR5}) is provided for each dubbing reflection, which is driven by rhythm comparison in a lenient way (\textbf{DR6}).

In this section, we initially present the user-interface of \system with a user walk-through in~\autoref{subsec:user_walkthrough}. Then we break down the system functions by introducing how the rhythm is extracted from any given speech in ~\autoref{subsec:rhythm_extraction}, how the rhythm is visualized on screen intuitively in ~\autoref{subsec:rhythm_visualization}, and how the rhythm comparison between two speech is achieved with tolerance in ~\autoref{subsec:rhythm_comparison} sequentially.

\begin{figure*}[ht]
    \centering
    \includegraphics[width=1\linewidth]{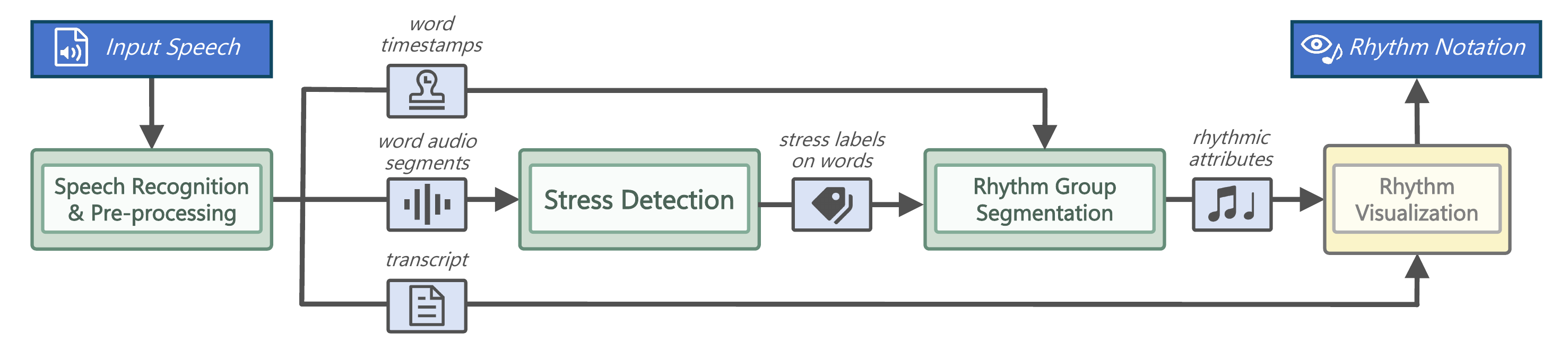}
    \caption{The \system pipeline consists of three main modules: (1) \textit{Speech Recognition and Pre-processing}, which transcribes speech into transcript with word-level timestamps and then segments audio into individual word clips; (2) \textit{Stress Detection}, which processes word audio segments to predict stress labels reflecting contextual and prosodic shifts in natural speech; and (3) \textit{Rhythm Group Segmentation}, which detects temporal regularities in stress intervals and segments them into groups accordingly. The resulting rhythmic attributes, including word-level stress labels, timestamps, and rhythm groups, are passed to the \textit{Rhythm Visualization} module, where speech rhythm is represented visually as rhythm notation. }
    \label{fig:pipeline}
\end{figure*}

\subsection{User Walk-through}
\label{subsec:user_walkthrough}

We introduce Leo, a first-year PhD student as well as a fourteen-year ESL learner, starting his research journey in an English speaking environment. Although he is able to communicate in English spontaneously without much concern in pronunciation, Leo still aims for making his English sound more natural, especially in terms of speech rhythm. Here, we present how he utilizes \system to understand, practice, and ultimately improve his English speech rhythm. 

As shown in ~\autoref{fig:interface}, the left column contains two basic components in all video dubbing applications. Leo can start with any English-speaking audiovisual materials that he enjoys imitating, and watch the scene from the video player (\autoref{fig:interface}: a). The target speech has been clipped into sentences automatically in advance and presented in a list (\autoref{fig:interface}: b). Leo is going to revoice the speech of the original speaker piece by piece. 

When it comes to practicing each dubbing clip, Leo should pay the most attention to the dubbing area (\autoref{fig:interface}: c), where the three learning stages (\textit{i.e.}, listening, repeating, and reflection) are integrated. 
The current learning stage is indicated by c0, where \textit{``Practice''} includes the listening and repeating stages while \textit{``Reflection''} is only for the reflection. Throughout the three stages, the target rhythm notation including the annotated script and corresponding dots along a horizontal timeline in c1 (\autoref{fig:interface}) remains on the screen visualizing the rhythm of target speech. Other components (\autoref{fig:interface}: c3-c5) appear only after Leo completes one dubbing attempt in the repeating stage and automatically moves to comparative reflection\textcolor{revision_color}{. 
We will} see how Leo goes through three stages.

\subsubsection*{\textbf{S1: Visual-aided Listening}}
By clicking the play button adjacent to the target speech script in c1, Leo initiates playback of the current dubbing clip's audio. To enhance speech rhythm perception, the script text and corresponding dots progressively appear from left to right in perfect synchronization with the audio playback.
While receiving authentic auditory input, Leo can simultaneously observe salient rhythm indicators.
First, word stress/de-stress is visually encoded through text font styles and dot fill patterns. Second, speech timing is represented by dots' horizontal positioning. What's more, stress interval regularities are highlighted with non-gray hues. Through this visual aids, Leo gradually gets familiar with the rhythm in current speech.

\subsubsection*{\textbf{S2: Visual-guided Repeating}} An additional record button is available in c1 in the former stage, allowing Leo to start his repetition when ready and move into the repeating stage.
During repeating, the annotated text and dots remain static on screen. It gives Leo a chance to anticipate upcoming rhythm patterns and thus make better preparation. 
Apart from that, a dynamic progress bar moves through the dots along the timeline, marking which words have been uttered in the target speech synchronously. 
They work together as visual cues, providing real-time rhythmic guidance for Leo to adapt to the original rhythm.

\subsubsection*{\textbf{S3: Comparative Reflection}} 
Upon completing the dubbing, the interface automatically transitions to the reflection view. 
The user rhythm notation in c2 faithfully records the rhythm of Leo's last dubbing, providing a visual evidence of his rhythm performance. 
% It shares a similar but vertically flipped layout with c1. 
By clicking the earphone button next to Leo's transcript, the latest audio recording can be played for Leo's review.
But he can also examine his rhythm in a more convenient way.
Facilitated by the parallel layout of c1 and c2, the horizontal offsets of corresponding words are intuitive to observe, which makes the temporal deviations apparent visually.
Besides, the waterfall-like component (\autoref{fig:interface}: c3) between c1 and c2 highlights each area where the stress intervals temporarily go consistent in the target speech. 
Leo further knows how his rhythm deviates from the target speech in those highlighted areas by clicking c3 for corrective feedback (\autoref{fig:interface}: c4). Contextual audio replay is available via the local replayer, two embedded playback controls in c4.
Moreover, words corresponding to the highlighted area will be covered by an curly brace (\autoref{fig:interface}: c5) under the Leo's transcript.
Its \textcolor{revision_color}{red color} indicates \textcolor{revision_color}{poor rhythm performance} right in this area. \textcolor{revision_color}{And it will turn green along with the feedback window once the desired rhythm is achieved.
}

Guided by the comparative analysis in the reflection stage, Leo returns to the listening stage and aims to implement corrective adjustments in his next dubbing attempt.

\subsection{Rhythm Extraction}
\label{subsec:rhythm_extraction}

English speech rhythm is determined by the timing of stressed syllables, creating variations in prominence over time. Given that, in English, each word contains either one primary stressed syllable or no stressed syllable at all, we can detect rhythmic strength in speech by identifying whether a word carries any stressed syllable.
Starting with the stress at the word level, by finding the regularities of stress intervals, the rhythmic patterns of English speech can be interpreted consequently.
Based on this, we propose an automated pipeline in \autoref{fig:pipeline} to extract the rhythm from any English speech into a special data structure ready for visualization. 

\subsubsection*{\textbf{Pipeline Overview}}
The pipeline takes speech audio as input and comprises of three modules, namely speech recognition and preprocessing, stress detection, and rhythm group segmentation. It produces all the necessary data for rhythm visualization, including the word-level timestamps and stress labels of rhythm notes and speech transcript. This pipeline serves for dual purposes in \textit{RhythmTA}. One is for dubbing material preparation which provides the rhythm of target speech as ground truths for dubbing. 
The other is user rhythm detection 
% instantaneously 
after each dubbing attempt. 
We then introduce the three modules one by one to provide a complete overview of the rhythm extraction pipeline.

\subsubsection*{\textbf{M1: Speech Recognition and Preprocessing}}
The first module aims to recognize the speech and segment every word based on the start and end timestamps. Specifically, we employ a lightweight offline automatic speech recognition model, VOSK\footnote{https://alphacephei.com/vosk/}. It transcribe speech with timestamps at the word level in a negligible delay.
For dubbing material preparation, if the speech is too long for a single take (empirically, more than 18 words), GPT-4o will be used to segment the transcript semantically into shorter sentences, typically ranging from 5 to 18 words.
Then based on the timestamps on words, the speech audio is sliced into a sequence of word audio segments for stress detection in the next module.

\subsubsection*{\textbf{M2: Stress Detection}}
This module takes word audio segments as input and detects whether a word was pronounced with any syllable stressed. It does not strictly distinguish lexical words (\textit{i.e.}, words that typically carry primary stress, as listed in dictionaries), because in natural speech, English speakers may stress functional words (e.g., `the,' `of,' `and') for emphasis or de-stress lexical words when they are less important. 
Since the module works directly on audio, it should capture these contextual stress shifts, as well as any deviant stress patterns, as they actually occur in speech.

Specifically, we first utilized a pre-trained wav2vec 2.0 representation model\footnote{https://huggingface.co/facebook/wav2vec2-base-960h} to extract features from raw audio. The detection model is then built using a Conformer architecture~\cite{gulati2020conformer} with two Conformer blocks. The model takes the wav2vec 2.0 representations as input and produces binary classification outputs indicating whether a word segment contains any stressed syllable. For model training, we utilized Aix-MARSEC dataset~\cite{auran2004aix}, which comprises approximately six hours of BBC radio broadcasts in British English with expert-annotated prosodic labels. The dataset provides syllable-level stress annotations, from which we derived word-level labels by checking whether any syllable in a word was stressed. The data were split into independent training, validation, and test sets.
On the test set, our model achieved a stress detection accuracy of \textbf{85.44\%}.

\subsubsection*{\textbf{M3: Rhythm Group Segmentation}}
\label{rhythm group segmentation}
This module detects temporal regularities in stress timing, where stress intervals become consistent, forming temporarily stable rhythmic beats at a particular pace.
First, we compute the intervals of consecutive stressed words. For the i-th stress interval $I_i$, we define it as:
$$
I_i = \frac{s_{i+1}+e_{i+1}}{2} - \frac{s_i+e_i}{2}
% I_i = (S_{i+1}+E_{i+1})/2 - (S_i+E_i)/2
$$
where $s_i$ and $e_i$ are the start and end times of the i-th word in speech respectively. 
Given a stress interval sequence $I=\{I_1, I_2, ..., I_{N-1}\}$ where $N$ is the number of stressed words, we design a sliding window algorithm to segment neighboring stress intervals with similar duration. 
We use the normalized Pairwise Variability Index (nPVI), proposed by Grabe et al. ~\cite{grabe2002durational} to measure local timing variability between successive speech units.
$$
nPVI = 100\left[\sum_{k=1}^{M-1}{\left|\frac{d_k-d_{k+1}}{d_k + d_{k+1}}\right|/(M-1)}\right]
$$

% \textcolor{red}{With similar motivation, Grabe and Low (2002) employed a measure of local timing variability originally developed by Francis nolan, the Pairwise Variability index, or PVi, that quantifies the degree to which successive units (often, but not necessarily, syllables) differ in duration. it has a normalized form, that uses the average interval length within each pair as a normalization factor:}

As detailed in \autoref{Alg: segmentation}, the sliding window starts at the first interval $I_1$ and maintains a list of candidate intervals for the current segment. It iteratively checks each subsequent interval and adds it to the list if the local nPVI remains below a threshold $\tau$ (empirically set to 18). If adding the next interval would exceed $\tau$, a new segment is created.

\begin{algorithm}
  \caption{Sliding-window Algorithm}
  \label{Alg: segmentation}
    \KwIn{Stress interval sequence $I$;
    nPVI threshold $\tau$.}
    \KwOut{Segmented stress intervals $G$.}  
    \BlankLine
    Initialize an empty segment $s$;

    \ForEach{$I_i \in I$}{
        Append $I_i$ to segment $s$;
        
        Compute $\alpha \leftarrow$ nPVI of segment $s$;
        
        \If{$|s| > 1$ and $\alpha > \tau$}{
            Remove $I_i$ from segment $s$;
            
            Append segment $s$ to set $G$;
            
            Update segment $s\leftarrow [I_i]$;
        }
    }
    Append segment $s$ to set $G$;
    
    \Return $G$;
\end{algorithm}

From the segmented stress intervals, we exclude segments with only one interval and retain those containing at least three stressed words as rhythm groups. Each rhythm group indicates that a steady beat occurs in the corresponding utterance, revealing important rhythmic patterns in the speech. Using the stress labels, rhythm group information, and speech transcript, we are able to visualize the speech rhythm with the following novel visual designs.

% With all the segmented stress intervals, we filter the segments with single interval out and keep those covering at least three stressed words as rhythm groups. Any rhythm group indicates steady beats happens in the corresponding utterance, providing the important rhythmic patterns in speech.

% With the stress labels, grouping information, as well as speech transcript obtained, the speech rhythm is ready to be visualized. 
%有了G怎么变成group xxxx

% Although there is no strict isochrony throughout a speech, or even a sentence, we surprisingly observed that there exists local stability of stress intervals frequently. [Fig] We call this segment including more than one stress interval a rhythm group. It means that the rhythm becomes steady temporarily right within this area.

\subsection{Rhythm Visualization}
\label{subsec:rhythm_visualization}
\begin{figure}[t!]
    \centering
    \includegraphics[width=1\linewidth]{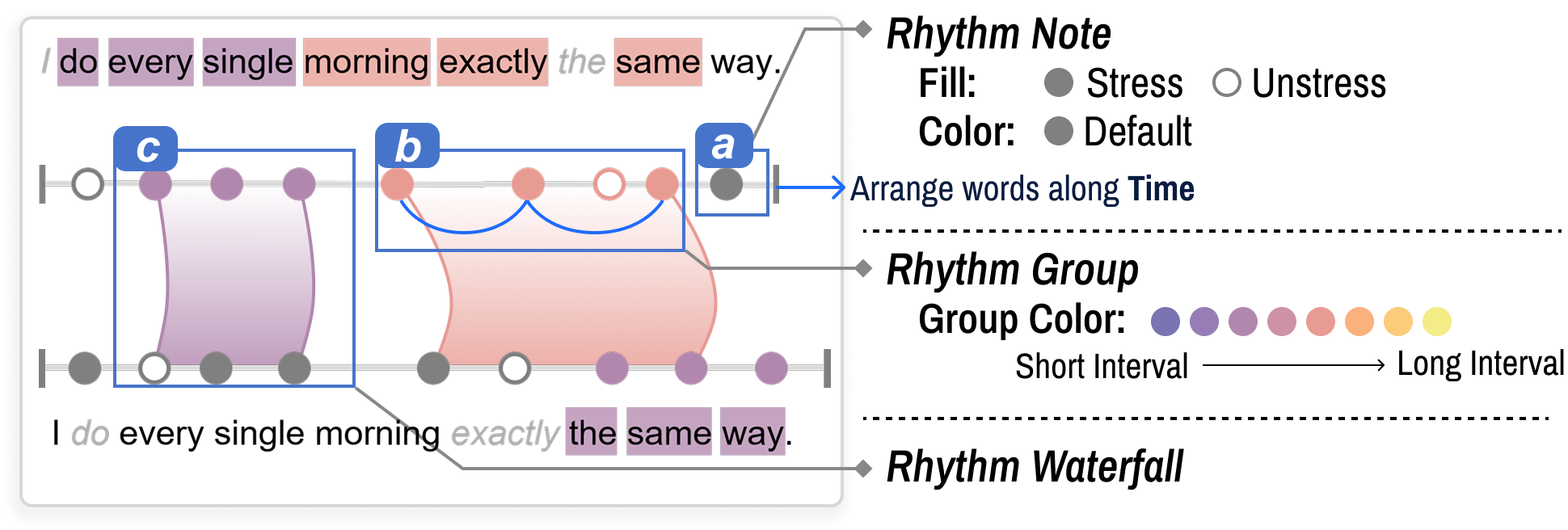}
    \caption{This figure illustrates how \system visualizes speech rhythm through three key components: (a) rhythm note, which shows the timing and stress of each word along a timeline; (b) rhythm group, which groups two or more consecutive stressed words with similar intervals and color-codes them based on interval length; and (c) rhythm waterfall, which visualizes the correspondence between rhythm groups in the target and user speech. }
    \label{fig:rhythm_vis}
\end{figure}

We propose a set of novel visual designs for rhythm visualization, as shown in~\autoref{fig:rhythm_vis}. 
Given the transcript and rhythmic attributes of a speech, we represent it as a rhythm notation. 
Our rhythm notation employs a dual-track layout to balance spatial closeness and individual readability between transcript words and rhythm notes.
Within the rhythm notes, rhythm groups are color-coded to highlight regularities of stress timing. 
Additionally, we introduce rhythm waterfalls to facilitate comparisons between two rhythm notations.
We introduce each design in details.

\subsubsection*{\textbf{Dual-track Layout}}
When visualizing speech, there are typically two layout options to consider: one that follows auditory information linearly over time, and another that aligns with the script, which is usually optimized for character-based readability. Speech rhythm is inherently time-linear, whereas a character-linear script offers the most comfortable reading experience for dubbing. To accommodate both, we adopt a dual-track layout, where a timeline runs parallel beneath the script. Subsequent rhythm notation designs will be based solely on this horizontal timeline.

\subsubsection*{\textbf{\note}}
Rhythm notes are the most basic elements of rhythm here, each representing a single word in speech. It is displayed as a circle whose horizontal position aligns with the center timestamp of its utterance. An encoding of being either filled or hollow is applied to the circle, indicating whether the corresponding word was stressed in speech. By default, all the circles are grey.

\subsubsection*{\textbf{\group}} In Section~\ref{rhythm group segmentation}, words which maintain steady beats with their neighbors can be segmented from speech. It indicates a periodicity of stress intervals locally, creating a prominent sense of rhythm. To visualize such typical utterances, any word within a rhythm group will follow an additional color encoding. Specifically, the same color will be assigned to all members of the group. The color is determined by the average stress interval: the longest intervals ($\geq$ 1.0 second, which rarely occur empirically) are represented by yellow, while the shortest intervals ($\leq$ 0.1 second, which are also rare) are represented by purple. These colors primarily serve to distinguish different rhythm groups, and represent the pace of rhythmic beats as well.

\subsubsection*{\textbf{\view}} It is designed for a parallel comparative view, where the rhythm notations of target speech and user speech are aligned to the left to reveal the discrepancies of stresses in time. A waterfall-like design starts from every rhythm group in the target notation respectively, falling down to cover the corresponding range of the user utterance. It serves to highlight the rhythmic prominent areas in target speech and anchor the local dubbing from users for deviation detection.

\subsection{Rhythm Comparison}
\label{subsec:rhythm_comparison}
We provide explicit feedback in addition to the intuitive visualization to further encourage user reflection. 
First, the word correspondence between the target speech and user speech is calculated using a fuzzy-matching Levenshtein algorithm. 
Then deviations from the target speech are calculated across all rhythm groups, including stress accuracy, beat stability, and pace similarity. 
Finally, corrective feedback is generated using rule-based methods while avoiding overly strict corrections.

\subsubsection*{\textbf{Word Mapping}}
Given a pair of transcripts from target speech and user speech, we use the Levenshtein distance algorithm to establish word-level alignments. 
Four alignment types are defined: match, replacement, deletion, and insertion. 
A match is identified if words from the target and user speech satisfy a fuzzy-matching threshold (threshold=0.62, computed via Python's difflib~\footnote{https://docs.python.org/3/library/difflib.html} library and determined empirically). 
A deletion occurs when the user omits a word present in the target speech, while an insertion represents an extra word in the user speech.

\subsubsection*{\textbf{Rhythmic Deviation Calculation}}
We apply a rhythmic deviation calculation method to words covered by rhythm waterfalls, aiming to examine whether the user's rhythmic prominence (induced by stress interval regularities) resembles that of the target speech. 
Specifically, for each rhythm waterfall, we calculate deviations in three aspects:
(1) \textbf{Stress accuracy} ($\bm{D}_{stress}$): the count of mismatched stresses/non-stresses in the matched and replaced words; (2) \textbf{Beat stability} ($\bm{D}_{beat}$): the count of words excluded from user rhythm groups; (3) \textbf{Pace similarity} ($\bm{D}_{pace}$): the range of color gradients covered by the absolute deviation between the target and user's average stress intervals.

\subsubsection*{\textbf{Tolerant Corrective Feedback}}
We provide corrective feedback for each rhythm waterfall in a lenient way.
To smooth the deviations counted in three dimensions, we apply two filtering steps. First, if $\bm{D}_{stress}>1$, set $\bm{D}_{beat}=\min(\bm{D}_{stress}, \bm{D}_{beat})$ to avoid duplicate deviation counts from misplaced stresses. Second, we sort deviations in descending order and reduce the smallest by 1 to prioritize major issues. Feedback is then assembled in descending order of deviations. For example, given $\bm{D}_{stress}=2, \bm{D}_{pace}=1, \bm{D}_{stress}=0$, the output would be:

\textit{``Do not forget to stress at [words]. You could speak a little faster.''}

If fewer than two deviations remain, the system displays a green positive assessment; otherwise, it turns red. For quick performance overviews, curly braces mark users' words covered by the current rhythm waterfall.

% \subsection{Pipeline Overview}
% \yanna{Pipeline is vague, try to change to other words like: Automatic Rhythm Detection?
% I suggests to seperate the detection methods with the system design. Then, the detection methods has its own pipeline in one subsection, the system design~~~ For this two part, start with an overview introduce what components will have and then go to the details in subsubsection.

% After introducing this section. Then it will be smoother to use a usage scenario to show  how to use our system by showing/going through all features and what effect we can achieve..
% Usage scenairos is a faked one, with the aim to show all features and show the perfect status you expected for you system.
% }

% \subsection{Rhythm Notation Strategy}

\subsection{Implementation}
% \subsubsection{System}
\system is a web-based program implemented in the React\footnote{https://react.dev/} framework. The audio and speech interface can be supported by the built-in speaker and microphone of a Lenovo ThinkPad T490s. A back-end server built in Python with Flask\footnote{https://flask.palletsprojects.com/en/stable/} is hosted remotely as well. 
We use VOSK's English model (vosk-model-en-us-0.22) for speech recognition, and perform stress detection on an NVIDIA RTX 4090 GPU.

\section{USER EVALUATION}
We conducted a within-subjects user study with twelve ESL learners to validate \textit{RhythmTA} 
\textcolor{revision_color}{
in terms of system usability, learning facilitation, learning improvement, and learning experience, compared to a baseline system simulating common dubbing applications.
}

\begin{figure*}[ht]
    \centering
    \includegraphics[width=1\linewidth]{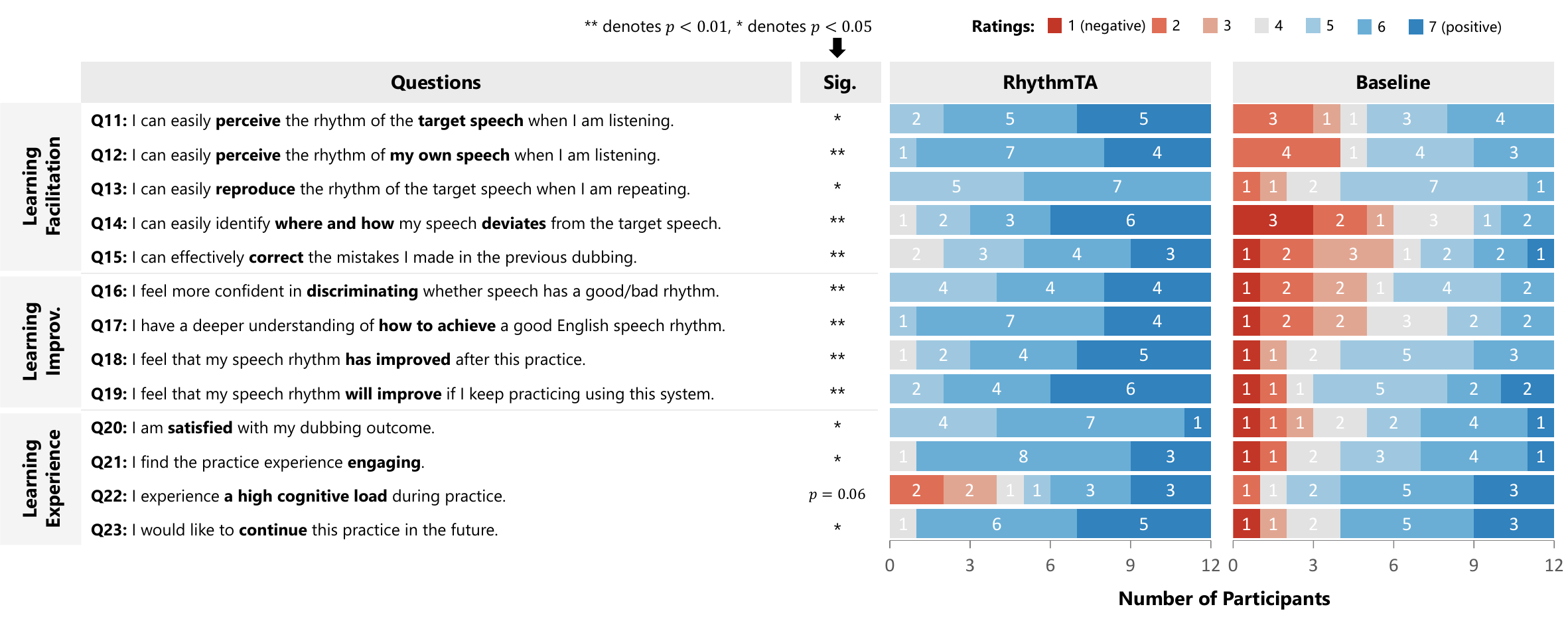}
    \caption{This figure displays user ratings for  \system and the baseline across \textcolor{revision_color}{learning facilitation (Q11-Q15), learning improvement (Q16-Q19),} and learning experience (Q20-Q23), measured on a 7-point Likert scale. Significant difference are marked with * ($p < 0.05$) and ** ($p < 0.01$). For consistency, ratings for the negatively worded question (Q22) have been reversed.}
    \label{fig:questionnaire}
\end{figure*}

\subsection{Study Setup} 

\subsubsection*{\textbf{Participants}}
We recruited twelve participants (six females, six males; aged from 21 to 31, $M =25.3 \pm 3.0$), all of whom are non-native English speaker with an average $16.1 \pm 4.1$ years of English learning experience. 
Participants' first languages included Chinese (Mandarin and Cantonese), Korean, Japanese, and German. 
Their English speaking proficiency, as measured by the IELTS speaking band, ranged from 6 (intermediate) to 8 (advanced) (M = 6.9 ± 0.6). For six participants, their scores were converted from equivalent speaking band scores in TOEFL, HKDSE, and CEFR.
% , sharing a profile of our target users: ESL learners who are able to make spontaneous speech and have little concern in pronunciation.
% \yanna{the profile of our target user is??? who can ake spontaneous speech and have little concern in pronunciation???? it is inappriate. 
% Moreover, why 6-8 are those learners with litter concern in pronunciation?? Strange}

% \subsubsection{Baseline}
% To build a baseline, we removed all the visual aids and feedback related to rhythm from \textit{RhythmTA} and kept the transcript of user speech alone in the reflection stage. 
% It simulates current dubbing applications with pronunciation reference only.

\subsubsection*{\textbf{Materials}}
We selected three video clips (V1-V3) from YouTube~\footnote{https://www.youtube.com/}, which are widely used in commercial dubbing applications such Lingodub. V1 was a 21-second conversation from an American TV comedy. It was used as a toy material only during the tutorial session to familiarize participants with the system interface and interaction flow. For the materials used in the formal study session, V2 and V3, we searched with the keyword \textcolor{revision_color}{``}university commencement address\textcolor{revision_color}{''} and found two 35-second clips from two public speech respectively.
To make the task manageable, we selected 8 continuous sentences for each of two formal video clips, aiming to maintain similar speech characteristics (e.g., number of words, speech rate, number of rhythm groups, and vocabulary difficulty).
Specifically, V2 contains 86 words, 11 rhythm groups, and an average speech rate of 147.4 word per minute, while V3 contains 74 words, 11 rhythm groups, and an average speech rate of 126.9 word per minute.

\subsubsection*{\textbf{Task}} 
Each participant completed two dubbing tasks using \system and baseline, respectively.
The baseline system is a simplified version of \textit{RhythmTA}, retaining only the transcript of user speech to simulate current applications that provide pronunciation feedback only.
%by removing all visual aids and feedback related to rhythm, but kept the transcript of user speech to simulate current application that provide pronunciation feedback only.
For each dubbing task,
% a video material with 8 sentences is provided. 
participants were given a video clip and required to finish dubbing in a sentence-by-sentence manner.
% For each sentence, the participant was encouraged to follow the listening-repeating-reflection looping, until feeling satisfied with the dubbing outcome.
They practiced each sentence repeatedly until satisfied with their performance, then proceeded to the next sentence.
We counterbalanced both (a) the assignment of video clips to systems and (b) the order in which the systems were used.

\subsubsection*{\textbf{Procedure}}
The user study lasted around 1.5 hours and was conducted through one-on-one, face-to-face meetings.
After obtaining participants' consent to record the session and collect their feedback for research purposes, the user study started with an introduction of English speech rhythm and overall procedures.
% and how to practice it through each system.
Then participants used the two systems successively. 
For each system, they first took a tutorial session using the toy material (V1) to familiarize themselves with the task, and system's components and functionality.
Once they felt comfortable using the system, they proceeded to complete the formal dubbing tasks with the assigned video clip (V2, V3).
After finishing the dubbing tasks for each system, participants filled out a questionnaire in a think-aloud protocol. 
The study concluded with a semi-structured interview where the participants shared their comments on the advantages and disadvantages of \system and its effect on their understanding of English speech rhythm.
During a preliminary pilot study with two non-experimental participants (PP1, PP2), we observed that participants felt anxious and reluctant to listen to their own recorded audio when others were present.
To reduce this discomfort,
% Thus, to reduce the unease of participants when speaking and replaying their audio in front of people, 
the study conductors left the room during two formal dubbing sessions.

\subsubsection*{\textbf{Questionnaire}}
\textcolor{revision_color}{
The questionnaire consists of 26 items on a 7-point Likert scale. 
The first 23 questions were identical for both systems, organized into four metrics: \textit{System Usability}, \textit{Learning Facilitation}, \textit{Learning Improvement}, and \textit{Learning Experience}.
For the first metric, we adopted the System Usability Scale (SUS)~\cite{bangor2008empirical} in Q1-Q10. 
The remaining three metrics examined all six design requirements in \autoref{subsec: design requirement}, with corresponding questions listed in \autoref{fig:questionnaire}. 
Specifically, the overall learning improvement (Q16-Q19) and experience (Q20-Q23) of the imitation-based dubbing practice validated DR1, while specific items (Q11-Q15, Q22) assessed DR2-DR6.
The final three items (Q24-Q26) particularly evaluated the intuitiveness
and helpfulness of three components unique to \textit{RhythmTA}: rhythm notes and groups, rhythm waterfalls, and the local replayer.
}

%edited
\subsection{Quantitative Analysis}
\textcolor{revision_color}{
% To evaluate \system via a within-subjects user study, we simulated general dubbing practice in the baseline system, which also typically involves listening, repeating, and reflection stages in common dubbing applications \cite{Lingodub, Liulishuo}. 
In this section, we first examined the time participants spent on practice iterations in different systems in Sec. \ref{subsubsec: time commitment}, and then analyzed the questionnaire responses in Sec. \ref{subsubsec: questionnaire}.
}

\subsubsection{\textbf{Time Commitment}}
\label{subsubsec: time commitment}
\textcolor{revision_color}{
To evaluate \system via a within-subjects user study, we simulated general dubbing practice in the baseline system, which also typically involves listening, repeating, and reflection stages in common dubbing applications \cite{Lingodub, Liulishuo}. 
In both systems, when users make a dubbing attempt, the mode automatically switches from \textit{``Practice''} to \textit{``Reflection''} (\autoref{fig:interface}) after they complete an uninterrupted recording. 
% Specifically, the system starts recording upon the user's click, automatically stops when the duration matches the target speech with a buffer time appended (1.2 second empirically), and submits the audio to the backend. 
This ensured participants consistently progress through listening, repeating, and reflection stages in every dubbing attempt.
% , driven by automated mode transitions.
In \textit{RhythmTA}, the visual-aided interaction in the first two stages is synchronized to video playback, incurring no additional time cost. In the final stage, \system supports rhythm comparison through interactive visualization, while the baseline requires replaying both speeches to detect differences, both of which need time for reflection.
}

\textcolor{revision_color}{
To investigate the actual time usage in both systems, we collected the following data from the user study: (1) the duration per dubbing attempt, and (2) the number of attempts per dubbing clip. 
We then used a paired t-test for significance analysis. The average time (in seconds) to complete one dubbing attempt (from entering \textit{``Practice''} mode to exiting \textit{``Reflection''} mode) was not significantly different between the two systems ($M = 27.87 \pm 5.61$ vs. $M = 26.03 \pm 4.14$; $Z = 1.27$, $p = .23$), suggesting \system introduced no notable time overhead across three stages. However, participants made more attempts per clip in \system ($M = 3.08 \pm 1.19$ vs. $M = 2.26 \pm 1.19$; $Z = 2.79$, $p < .05$), indicating greater willingness to iterate.
}

\subsubsection{\textbf{Questionnaire Results}}
\label{subsubsec: questionnaire}
\textcolor{revision_color}{
We utilized the Wilcoxon signed-rank test~\cite{wilcoxon1963critical} to analyze the
significance of questionnaire responses in four metrics: \textit{System Usability}, \textit{Learning Facilitation}, \textit{Learning Improvement}, and \textit{Learning Experience}. 
Results for the latter three metrics are shown in \autoref{fig:questionnaire}.
% \textcolor{red}{All six design requirements in \autoref{subsec: design requirement} were examined. 
% The overall improvement (Q16-Q19) and experience (Q20-Q23) of the imitation-based dubbing practice validated \textbf{DR1}, while specific questionnaire items (Q11-Q15, Q22) assessed \textbf{DR2-DR6}.}
We also investigated the intuitiveness and helpfulness of three key components in \textit{RhythmTA}.
}

\begin{itemize} [leftmargin=*]
\item \textbf{\textit{System Usability (Q1-Q10)}}. 
We assessed system usability using the System
Usability Scale (SUS)~\cite{bangor2008empirical}.
\system received a SUS score of 82.78, while the baseline received 83.47.
Both scores fall within the ``excellent'' range and indicate usability levels exceeding approximately 90\% of applications~\cite{bangor2008empirical, sauro2016quantifying}. 
There was no significant difference between the two systems in most SUS questions, except for Q1 (\textit{``I think that I would like to use this system frequently''}) and Q5 (\textit{``I found the various functions in this system were well integrated''}). 
Participants rated \system as more favorable (Q1: $M=5.67\pm 1.07$ vs. $M=4.83\pm 1.70$; $Z=2.46, p<.05$), and better integrated in terms of functionality (Q5: $M=6.25 \pm 0.62$ vs. $M=5.25\pm1.55$; $Z=2.16, p<.05$) than the baseline.
These ratings demonstrate that, although \system offers more features, it strikes a balance between complexity and usability, as reflected in high ratings on positively framed questions ($M = 6.04\pm 0.27$) and low ratings on negatively framed questions ($M = 2.11\pm 0.37$).

\item \textcolor{revision_color}{\textbf{\textit{Learning Facilitation (Q11-Q15)}}. 
We compared the learning facilitation provided by the two systems, by examining DR2-DR5 respectively. \system demonstrated significantly better performance across all related measures.
% - DR2 (Q11&12): Rhythm perception
% - DR3 (Q13): Guidance for reproducing rhythm
% - DR4 (Q14): Comparison for identifying deviations
% - DR5 (Q15): Corrective feedback
First, for rhythm perception (\textbf{DR2}), which \textit{RhythmTA} addressed through visual aids, participants found it easier to perceive rhythm in \system than in the baseline, 
both when listening to the target speech (Q11: $M = 6.25 \pm 0.75$ vs. $M = 4.33 \pm 1.67$; $Z = 2.57$, $p < .05$) and their own recordings (Q12: $M = 6.25 \pm 0.62$ vs. $M = 4.17 \pm 1.70$; $Z = 2.83$, $p < .01$). 
Second, for rhythmic guidance during dubbing (\textbf{DR3}), \system enabled easier rhythm reproduction compared to the baseline (Q13: $M = 5.58 \pm 0.52$ vs. $M = 4.50 \pm 1.09$; $Z = 2.49$, $p < .05$).
Third, \system also showed greater facilitation in speech rhythm comparison (\textbf{DR4}), helping participants identify rhythmic deviations more intuitively (Q14: $M = 6.17 \pm 1.03$ vs. $M = 3.25 \pm 1.87$; $Z = 2.81$, $p < .01$).
Finally, participants also corrected mistakes (\textbf{DR5}) more effectively in \system (Q15: $M = 5.67 \pm 1.07$ vs. $M = 3.92 \pm 1.88$; $Z = 2.85$, $p < .01$).
}

% For \textbf{rhythm production (Q13-Q15)}, participants found it easier to reproduce the rhythm ($M = 5.58 \pm 0.52$ vs. $M = 4.50 \pm 1.09$; $Z = 2.49$, $p < .05$), identify rhythmic deviations ($M = 6.17 \pm 1.03$ vs. $M = 3.25 \pm 1.87$; $Z = 2.81$, $p < .01$), and correct mistakes ($M = 5.67 \pm 1.07$ vs. $M = 3.92 \pm 1.88$; $Z = 2.85$, $p < .01$) when imitating the target speech using \system compared with the baseline. 

\item \textcolor{revision_color}{\textbf{\textit{Learning Improvement (Q16-Q19)}}. 
}
For the self-reported rhythm improvement, \system significantly outperformed the baseline, suggesting pedagogical benefit in ESL learners' speech rhythm training.
Specifically, participants reported greater confidence in evaluating rhythm quality (Q16: $M = 6.00 \pm 0.85$ vs. $M = 3.92 \pm 1.68$; $Z = 2.84$, $p < .01$), a deeper understanding  of how to achieve a good rhythm (Q17: $M = 6.25 \pm 0.62$ vs. $M = 3.75 \pm 1.60$; $Z = 2.95$, $p < .01$), noticeable rhythm improvement after practice (Q18: $M = 6.08 \pm 1.00$ vs. $M = 4.58 \pm 1.44$; $Z = 2.84$, $p < .01$), and a stronger belief in continued improvement with further use (Q19: $M = 6.33 \pm 0.78$ vs. $M = 4.83 \pm 1.80$; $Z = 2.70$, $p < .01$). 

\item \textbf{\textit{Learning Experience (Q20-Q23)}}. 
We investigated the participants' learning experience and found an overall more positive response to \system than to the baseline.
Specifically, participants showed significantly higher satisfaction with the dubbing outcomes (Q20: $M = 5.75 \pm 0.62$ vs. $M = 4.58 \pm 1.83$; $Z = 2.23$, $p < .05$), greater engagement during practice (Q21: $M = 6.08 \pm 0.79$ vs. $M = 4.75 \pm 1.77$; $Z = 2.56$, $p < .05$), and a stronger willingness to continue using the \system in the future (Q23: $M = 6.25 \pm 0.87$ vs. $M = 5.25 \pm 1.87$; $Z = 2.23$, $p < .05$).
\system also showed slightly but not significantly higher perceived cognitive load than the baseline (Q22 flipped: $M = 4.83 \pm 1.95$ vs. $M = 5.58 \pm 1.44$; $Z = -1.90, p > .05$). 
\textcolor{revision_color}{This suggests that cognitive load during \system use remained under control, likely because the system employed a lenient rhythm assessment to minimize user anxiety (\textbf{DR6}).}

\item \textbf{\textit{Design Validation (Q24-Q26)}}. 
We evaluated the intuitiveness and helpfulness of three key components in \textit{RhythmTA}.
The design of rhythm notes and rhythm groups were rated as highly intuitive (Q24: $M = 6.33 \pm0.62$), and the design of  rhythm waterfalls received an even higher intuitiveness score  (Q25: $M = 6.58 \pm0.64$). 
Participants also found the local replayer helpful for reviewing specific segments of speech during self-reflection (Q26: $M=6.5 \pm0.87$).
\end{itemize}

\subsection{Qualitative Findings}
We present the qualitative findings about \system based on participants' feedback and behaviors in the study.

\subsubsection{\textbf{\system enables ESL learners to practice speech rhythm independently}}\textcolor{white}{.}

All participants completed the dubbing practices using \system without any external help.
% As statistics, participants 
They reported that \system improved their perception of speech rhythm, made it easier to compare their own speech with the target, and increased their awareness of self-adjustment for their speech rhythm.
For example, P3 recognized the help of \system in rhythm improvement, \textit{``It allowed me to learn how native speakers naturally modulate their speech by pauses and stress based on contexts—something I had noticed before but could never analyzed closely''}. 
P6 was also convinced that the system was the ideal tool for her needs, because \textit{``while I can figure out pitch and practice intonation by myself, I lack rhythmic awareness and would find this system highly valuable''}.
% As statistics, enhanced perception of rhythm, intuitive comparison between speech, and awareness of self-adjustment were all reported in the interviews (P1-P12).
% It 
This
reflected how \system supports independent practice by addressing the three challenges of ESL rhythm training (\textbf{C1}: limitations in rhythm discrimination, \textbf{C2}: difficulties in speech comparison, and \textbf{C3}: unconscious L1 interference). 

% \yanna{I think this part are not well supported by the user study, may need to put it on discussion about: Support independent english rhythm practice}

\subsubsection{\textbf{Rhythm notation enhances ESL learners' perception and understanding of speech rhythm}} \textcolor{white}{.}

\textbf{Rhythm notation compensates for ESL learners' rhythm discrimination limitations.} In \textit{RhythmTA}, rhythm notation visualizes the stresses and their timing patterns for all speech.
P3 described, \textit{``I cannot tell the rhythm either in the target speech or my own by ears, but I now can know it pretty well with the visual aids.''} Similarly, P2 remarked, \textit{``Learners who are less sensitive to spoken languages could get the most benefit from the system.''} 
% The facilitation provided by rhythm notation in different levels was mentioned by participants specifically as follows.
Specifically, some participants highlighted that rhythm notation helped them better perceive stresses or pauses in speech.
For example, P4 and P8 particularly noted their need to know not only the existence but also the durations of \textit{``breaks''} in speech had been satisfied.
\textit{``I can visually see how the native speaker pauses to emphasize ... I used to pause in my speech as well, but didn't know whether it was long enough until using this system''} (P8). 
P5 similarly appreciated how the fill style of rhythm notes provided clear stress information. 
Moreover, participants also appreciated that rhythm notation made temporal variations and regularities in speech apparent at a glance.
This helped them easily identify some aurally subtle traces of rhythm that were difficult to hear, such as speech pace (P2, P5, P11) and stress interval regularities (P6, P7, P10).
% , can be identified easily with rhythm notation.
It was probably because \textit{``the information linear to time is now unfolded in parallel''} (P6), where a duration of time was projected to length along a horizontal axis.
P12 described this refreshing experience as \textit{``nailing the rhythm down on the screen''}.
% It demonstrated that rhythm notation made the temporal variations and regularities in speech become apparent at a glance.

\textbf{Rhythm notation provides a lens for examining prominence in native speech.}
In \textit{RhythmTA}, rhythm notation highlights the stressed words and their corresponding notes as a group, once if stress timing becomes relatively isochronous.
Seven participants (P1, P5-P7, P9, P10, P12) expressed appreciation of rhythm groups in the interviews.
Acknowledging that rhythm groups merely disclosed speech prominence by steady beats, they interpreted it from different interesting perspectives. P9 and P10 shared a similar opinion that \textit{``rhythm groups trim the sentence into succinct chunks''}. P10 explained that it highlighted what words were essential information in the speech while others were not, which \textit{``... offered me a global overview of the sentence structure, and then I can allocate my attention appropriately''}. P6 tended to believe that the grouping was brought by proximity in semantics. While P8 held the same speculation with P6, he reported confusion when seeing this example \textit{``... every single [\textbf{morning exactly} the \textbf{same}] way''} where the bold words were grouped. P12, from another perspective, suggested that the rhythm groups usually appeared when the speaker showed a rather strong intention, noting \textit{``greater rhythmic regularity enhances speech force''}. The highlight of timing regularities gave participants opportunities to closely observe and interpret the rhythmic patterns in native English speech.

\subsubsection{\textbf{The parallel comparative view renders self-monitoring intuitive for ESL learners}}\textcolor{white}{.}

\textbf{The parallel view is praised for its intuitiveness in monitoring rhythm performance} by all participants. In this view, rhythm notations of the target speech and participants' speech were vertically parallelized and horizontally left-aligned. P3 noted that it made the comparison between speech possible by simply observing the x-offsets of corresponding notes, \textit{``... I can tell whether each word in my dubbing catches the timing of the target utterance easily.''} P6 commented on the parallel comparative view that it remarkably released the tedious labor and memory overhead from listening to two speeches repeatedly when she wanted to find the differences in the baseline.

\textbf{Rhythm waterfalls further facilitate deviation detection}. Besides checking the deviations in single words, nearly all participants (P1-P3, P5-P12) mentioned that they benefited from the design of rhythm waterfalls, which mapped each target rhythm group to its corresponding dubbing segment. P9 stated that he appreciated how the colorful waterfall-like visualization highlighted areas of interest concisely in his speech, so that \textit{``I do not need to analyze the rhythm by taking the sentence as a whole''}. P10 noted that rhythm waterfalls provided an anchoring effect: \textit{``the boundaries act as anchors, informing me of which text corresponds to each rhythm note without requiring hover-based queries''}. With the facilitation provided by rhythm waterfalls, eleven participants expressed that they could quickly identify their rhythmic deviations from the target speech. P12 commented: \textit{``It strikes a balance between being informative and being visually complex.''} However, one participant (P4) held an opposite opinion, indicating the discouragement caused by unmatched rhythm was intensified by rhythm waterfalls: \textit{``I prefer to keep only the parallel view but without the waterfalls.''}

\textbf{The local replayer is less frequently used}. 
Integrated into the corrective feedback window, the local replayer provides an auditory reference, allowing users to review specific segments of their speech. P10 stated: \textit{``The side-by-side layout of local relayer makes it convenient to switch between the target speech and my own for comparison.''} Yet, this component was less frequently used by participants as the authors observed from screen recordings. P6 shared her strategy of reflection which helped to explain: \textit{``I checked whether my rhythm went well directly based on the parallel view, and only used the local replayer if I saw severe deviations.''}

% offset, 上下对比 直观

\subsubsection{\textbf{The visual cues and corrective feedback ensure effective self-adjustment for ESL learners}}\textcolor{white}{.}

\textbf{Visual cues signal ESL learners to adjust speech rhythm during dubbing.}
Participants all reported the experience of how they modulated their voices according to the visual cues in the repeating stage. 
The cues, including rhythm notation and a progress bar, were recognized as helpful.
For example, P4 commented they \textit{``provide guidance on how to speak exactly like the speaker rhythmically''}.
P12 thought they \textit{``help to forecast the following paralinguitic information, reminding me of how I should produce my speech next''}). Specifically, two participants (P4, P8) were most interested in following the pausing after words, while six participants (P1, P3, P5, P6, P9, P11) indicated that they paid the most attention to adjusting the stress timing (P5: \textit{``when to stress the words''}). 
P9 stated he benefited from being lead by the progress bar which denoted the actual timing of the target speech. 
However, on the contrary, P3 and P6 tended to feel nervous once they failed to follow the original progress. 
P8 shared a similar cognitive load when trying to catch all target rhythm notes accurately, but he characterized it as \textit{``a necessary learning experience''}.

\textbf{Corrective feedback provides a clear goal in continuous dubbing practices.} 
As participants' speech rhythm were assessed in stress accuracy, beat stability, and pace similarity to the target speech, P1 appreciated the subsequent advice based on his rhythmic flaws, by noting \textit{``it presented me a specific direction to optimize''}. 
P9 mentioned that he got positive feelings whenever \system feedback appeared \textit{``almost instantly''} after his dubbing—\textit{``it satisfied my curiosity about whether my rhythm was good enough''}.
A common tendency of repeated dubbing attempts till the system feedback turned positive was observed among nine participants (P1-P5, P7-P9, P12).
Eight participants (P2, P3, P5, P6, P9-P12) remarked the feedback provided by \system as valid enough without any confusion or annoyance, while two participants (P1, P8) regarded it as helpful except for slight discrepancies as they perceived in stress detection.
A minority of participants (P4, P7) pointed out that the text advice was a little excessive for them to handle properly.

\section{DISCUSSION}
We discuss the following two topics about visual-aided speech rhythm training: (1) personalization of \textit{RhythmTA}, and 
\textcolor{revision_color}{(2) implications for ESL rhythm training. We conclude this section with limitations and future work of this research.
}

\subsection{Personalization of \system}
We discover the following potential directions for tailoring \system to more diverse user contexts.

\subsubsection*{\textbf{Modality Preference Adaptation}}
\system combines auditory and visual modalities, with visual aids designed to complement learners' deficiencies in auditory perception. 
During the user study, we observed distinct modality preference among participants.
Some participants prioritized visual feedback to identify deviation and referred to audio only when necessary, while some participants consistently relied on audio for confirmation, regardless of the visual feedback. 
This divergence may stem from differences in participants' perceptual sensitivity or inherent trust in certain modalities.
This indicates an opportunity for an adaptive interface that dynamically adjust the modality prominence based on user interactions to better support learners with various preference.

\subsubsection*{\textbf{Dubbing Material Customization}}
In our user study, participants expressed excitement when using clips from familiar TV shows (e.g., P2, P7), highlighting the potential of learner-selected materials to boost learning motivation.
\system can automatically extract and visualize rhythm for any English speech, providing ESL learners the freedom to choose dubbing materials from any media, covering any topic, speaker, or English accent.
It is essential to customize materials to align with learners’ interests in order to promote learner-driven practice and sustain long-term engagement.
\textcolor{revision_color}{
Moreover, speech rhythm varies by occasion and personal style. For example, someone who wants to sound like a comedian or a lecturer should probably adopt distinct rhythmic flows. 
With \textit{RhythmTA}, users can practice daily conversations, rehearse for specific speaking contexts, or even mimic a celebrity’s speaking style, simply by selecting imitation targets that match their goals.
With this flexibility, future work could explore clustering scenario-specific rhythmic patterns and recommending examples tailored to user needs (e.g., small talk, job interviews, and public speaking).
}

\subsubsection*{\textbf{Feedback and Tasks Configuration}}
In our user study, participants expressed strong interest in personalizing various aspects of the rhythm training experience.
For instance, P10, an advanced learner, suggested the flexibility to adjust dubbing length, preferring to practice with longer discourse-level materials rather than isolated sentences.
Other participants also noted a desire to selectively focus on certain aspects of corrective feedback, based on their own perceived weaknesses or learning goals.
These observations highlight the need for a learner model that can automatically adapt content complexity and feedback granularity to match the learner’s proficiency and goals.
% A promising direction is to incorporate a learner model that automatically adjusts training content and feedback granularity based on users’ performance history and interaction behavior.

\subsection{Implications for ESL Rhythm Training}
\textcolor{revision_color}{
We summarize the implications of \system for ESL rhythm training, aiming to inspire future research in broader contexts. 
Potential directions include automating causal diagnosis of rhythmic flaws, generalizing to more diverse L1 backgrounds, extending to spontaneous speech scenarios, and optimizing multimodal engagement. 
}

\subsubsection*{\textbf{Trade-off between Rhythm and Enunciation}}
\textcolor{revision_color}{
Rhythm involves modulating stress and its timing, often requiring less important words to be spoken more rapidly ~\cite{agata2024teaching}. 
This creates a tension between maintaining precise enunciation and achieving a faster pace in certain words to establish rhythm.
Native speakers usually resolve this conflict intuitively through connected speech phenomena (e.g., linking and vowel reduction via schwa sounds), which shorten intervals between stressed syllables~\cite{agata2024teaching}. 
We observed cases in the user study where participants rushed their speech to catch up with the target rhythm, particularly in words that required consonant-to-consonant linking.
This likely stems from insufficient mastery of connected speech techniques.
It aligns with E1’s approach, which he mentioned in the formative study: he usually introduces syllable- and word-level connected speech concepts to students before advancing to rhythm training.
Future work could focus on automatically diagnosing such causal factors at the syllabic or lexical level for rhythmic flaws in user speech and providing tailored tutorials.}

\subsubsection*{\textbf{Generalizability across L1s}}
\textcolor{revision_color}{
\system visualizes speech rhythm through stress and its timing, which are the core rhythmic features in English. It is designed to support a wide range of ESL learners whose L1s do not exhibit similar patterns in stress timing. Adjustments may be needed, however, since \system shows stress in a binary form of rhythm notes for visualization simplicity, without presenting how stress is acoustically realized (e.g., pitch, volume, and duration). If learners’ L1s share English’s stress-timing patterns but differ primarily in stress production, future systems should visualize prosodic details to accommodate these differences. This may require visualization designs that can incorporate richer information while keeping clarity~\cite{xie2020interactive}.
}

\subsubsection*{\textbf{Extension to Spontaneous Speech}}
\textcolor{revision_color}{
\system adopts dubbing practice, an imitation-based approach commonly used in ESL rhythm training. 
It evaluates user speech rhythm by comparing it to an ideal target. 
For spontaneous speech scenarios, however, which allow users to modulate their rhythm freely, 
realistic native speech synthesized from user transcripts via text-to-speech models \footnote{https://elevenlabs.io/} can serve as rhythmic ground truths in \textit{RhythmTA}. 
Notably, varying communicative intents may require different rhythmic patterns even for identical content.
Future work could therefore explore generating multiple rhythmic references to capture this flexibility, or fine-tuning audio-language models ~\cite{chu2024qwen2} for end-to-end rhythm evaluation by leveraging their semantic understanding capabilities.
}

\subsubsection*{\textbf{\textcolor{revision_color}{Balancing Multimodal Reliance}}}
\system integrates both visual and auditory modalities to support rhythm learning. 
Participants generally appreciated the helpfulness of visual aids, such as rhythm notes and rhythm groups, for perceiving rhythm, guiding reading, and facilitating comparison.
However, we also observed shifts in modality reliance in the study.
Several participants reported they stopped listening to the original audio once they trusted the visual feedback. 
This behavioral shift suggests a risk of learners becoming overly dependent on visual cues.
While this indicates high trust in the visual design, it also underscores the importance of maintaining a balance between visual and auditory engagement.
Since visualizations cannot capture all acoustic features, such as pitch and intonation, visual aids should complement rather than replace auditory perception.
Future designs may explore adaptive strategies that gently encourage re-engagement with auditory input or highlight when critical prosodic information may be missed visually.
\textcolor{revision_color}{
Beyond avoiding unbalanced engagement, the optimal selection of the assisting modality remains underexplored.
Extensive experiments could be conducted across all alternative modalities (e.g., vision, haptics, and audition) in the future to find the most effective and unobtrusive modality choice for rhythmic aids.
}

\subsection{Limitations and Future Work}
\system presents several limitations that suggest directions for future work. 
First, the current rhythm extraction models is trained on
the Aix-MARSEC~\cite{auran2004aix} dataset, which primarily features speakers with British English accents. 
This may limit its adaptability to the diverse accents of ESL learners in terms of stress detection. 
Given that prosodically annotated datasets are costly and scarce, future improvements should explore leveraging larger, more varied corpora if they become available.
Second, the system’s reliance on visual aids may limit accessibility for users with visual impairments or color vision deficiencies. 
To address this, future work should explore accessible visual design (e.g., colorblind-safe palettes~\cite{hau2015colorbless}) and introduce multimodal support, such as audio-based rhythm scaffolding~\cite{maculewicz2016investigation, cason2015bridging} or vibrotactile cues~\cite{holland2018haptics, liao2017dwell} delivered via wearable devices. 
Finally, our evaluation was limited to a single 90-minute session, capturing only participants’ initial impressions and short-term learning outcomes.
Longitudinal studies in real-world learning settings are needed to assess long-term effectiveness and sustained engagement.

\section{CONCLUSION}
\system offers ESL learners crucial assistance in independent speech rhythm practice, integrating support into three typical stages of imitation-based learning methods. 
A rhythm extraction pipeline and a set of visual designs represent English speech rhythm as intuitive, comparison-friendly visualizations. 
The system addresses three key challenges: limitations in rhythm discrimination, difficulties in speech comparison, and unconscious L1 (first language) interference. 
Through user study, we demonstrated that \system effectively enhances learners’ rhythm perception and has significant potential to improve rhythm production.

\section*{ACKNOWLEDGMENTS}
We thank the participants in formative and user studies for their meaningful support, and the anonymous reviewers for their constructive feedback. 
We are also grateful to Liwenhan Xie, Dr. Nick Wong, Yanjia Li, and Qi Wang for their valuable advice.
This paper is supported by the Quality Education Fund from Hong Kong Education Bureau under the project titled \textit{``AI-assisted Virtual Reality English Speaking Program for Secondary Students''}.
% \end{acks}

% \bibliographystyle{abbrv}
\bibliographystyle{ACM-Reference-Format}
\bibliography{References}

\end{document}